\documentstyle[prl,aps,twocolumn]{revtex}
\input epsf
\begin{document}

\twocolumn[\hsize\textwidth\columnwidth\hsize\csname@twocolumnfalse\endcsname

\title{Competing roughening mechanisms in strained heteroepitaxy: a
  fast kinetic Monte Carlo study
}

\author{Chi-Hang Lam$^{1}$, Chun-Kin Lee$^1$, and Leonard M. Sander$^2$} 

\address{
$^1$Department of Applied Physics, Hong Kong Polytechnic University,
Hung Hom, Hong Kong\\ $^2$Michigan Center for Theoretical Physics,
Department of Physics, Randall Laboratory,
University of Michigan, Ann Arbor, MI 48109-1120, USA\\ }
\date{\today}
\maketitle

\begin{abstract}
  We study the morphological evolution of strained heteroepitaxial
  films using kinetic Monte Carlo simulations in two dimensions. A
  novel Green's function approach, analogous to boundary integral
  methods, is used to calculate elastic energies efficiently. We
  observe island formation at low lattice misfit and high temperature
  that is consistent with the Asaro-Tiller-Grinfeld instability
  theory. At high misfit and low temperature, islands or pits form
  according to the nucleation theory of Tersoff and LeGoues.
\end{abstract}


]

Coherent three-dimensional (3D) islands in strained heteroepitaxial films
are of great interest because they can self-assemble as quantum dots 
for possible advanced
optoelectronic applications \cite{Shchukin,Politi}. They are observed
in a variety of film-substrate combinations including
Ge/Si, InAs/GaAs, InAs/InP, etc. In these systems, island formation
follows the Stranski-Krastanov mode. Initially, two-dimensional (2D)
layer-by-layer growth leads to a flat wetting layer under stress.
Beyond a threshold film thickness, 3D islands emerge
on top of the wetting layer partially relieving the stress.
The precise island formation mechanism is currently under intensive
debate.  According to the nucleation theory of Tersoff and
LeGoues, the growth of stable islands requires overcoming an energy
barrier associated with a critical island size \cite{Tersoff_LeGoues}.
However, experiments reveal gradual development of ripples
\cite{Floro} or pre-pyramids \cite{Vailionis} at the initial stage of
island formation. These observations are more consistent with the
Asaro-Tiller-Grinfeld (ATG) linear instability theory, which predicts
that morphological perturbations at sufficiently long wavelengths grow
spontaneously and steadily \cite{Asaro,Srolovitz}. It was originally
proposed for smooth surfaces, but extensions to faceted ones based on
non-equilibrium deposition conditions \cite{Tersoff01} or finite
vicinality of substrates \cite{Eisenberg} have been suggested.

To better understand the roughening mechanism of strained layers, we
have performed kinetic Monte Carlo simulations using an atomistic
model \cite{Orr,Barabasi,Khor,Meixner}.  This approach is
computationally very intensive but can reliably account for both
lattice discreteness and non-equilibrium conditions.  Previous
simulations successfully demonstrated island formation in strained
layers but only via the nucleation mechanism
\cite{Orr,Barabasi,Khor}. In this work we introduce significantly more
efficient algorithms. Thus we can explore a much wider range of conditions and
observe a rich variety of morphologies in better general agreement
with experiments. In particular, by lowering the lattice misfit and
raising the temperature, the roughening mechanism crosses over from
nucleation to instability controlled.

We adopt the 2D ball and spring model of heteroepitaxy
defined on a square lattice first studied by Orr et al \cite{Orr} and
subsequently by Barab\'{a}si \cite{Barabasi} and Khor and Das Sarma
\cite{Khor}.  Simulations in 3D limited to submonolayer coverage were
performed by Meixner et al \cite{Meixner}. Our model parameters are
appropriate to
 the widely studied Si$_{1-x}$Ge$_x$/Si system.  We assume a substrate
lattice constant $a_s=2.715$\AA ~ so that $a_s^3$ gives the correct
atomic volume in crystalline silicon.  The lattice constant $a_f$ of
the film material is related to the lattice misfit
$\epsilon=(a_f-a_s)/a_f$ which has a compositional dependence
$\epsilon=0.04 x$.
Nearest and next nearest neighboring atoms are directly connected by
elastic springs with force constants $k_N=13.85eV/a_s^2$ and
$k_{NN}=k_N/2$ respectively. This choice gives the correct modulus
$c_{11}$ of silicon and a shear modulus constant along tangential and
diagonal directions, despite a slight anisotropy in the Young's
modulus.  The elastic couplings of adatoms with the rest of the system
are weak and are completely neglected for better computational
efficiency.  Solid-on-solid conditions and atomic steps limited to at
most two atoms high are assumed.  Every topmost atom in the film can
hop to a random topmost site $s$ columns away where $s=\pm 1$, $\pm
2$, ...  or $\pm s_{max}$ with equal probability.  
Previous simulations allowed only nearest neighbor hopping (i.e.
$s_{max}=1$) \cite{Orr,Barabasi,Khor,Meixner}.  To speed up the
simulations, we put $s_{max}=8$ or 20 respectively for $x > 0.6$ or $x
\le 0.6$.  These hopping ranges are much shorter than the dimensions
of the relevant structures (islands or pits) on the films
and we have checked that decreasing $s_{max}$ does not alter our
results.  The hopping rate $\Gamma_m$ of a topmost atom $m$ follows an
Arrhenius form:
\begin{equation}
\label{rate}
\Gamma_m = 
{R_0}\exp \left[ -\frac{n_m \gamma  
- \Delta E_m - E_0}{k_{B}T}\right]
\end{equation}
Here, $n_m$ is the number of nearest and next nearest neighbors of
atom $m$. We choose a bond strength $\gamma=0.4eV$ which will be
explained later. The energy $\Delta E_m$ is the difference in the
strain energy $E_s$ of the whole lattice at mechanical equilibrium
when the site is occupied versus unoccupied.  Finally, we put
%$E_0=-(0.67eV-3\gamma)$ 
$E_0=0.53$eV and $R_0=2D_0/(\sigma_s a_s)^2$ with $D_0=5.2\times
10^{13}\mbox{\AA}^2 s^{-1}$ and 
$\sigma_s^2 = \frac{1}{6}(s_{max}+1)(2s_{max}+1)$.
This gives the appropriate adatom diffusion coefficient for silicon (100)
\cite{Savage}. Our model follows detailed balance.

The simulations involve intensive computations resulting solely from
the long-range nature of elastic interactions. Practically all the CPU
time is spent on the repeated calculations of $E_s$ which is needed to
find $\Delta E_m$ and hence $\Gamma_m$ in Eq.  (\ref{rate}).  The
elastic problem is formulated as follows.  First, a flat film is
homogeneously strained \cite{Politi}.  This provides a convenient
reference position with displacement $\vec{u}_i=0$ for every atom $i$.
In general, the elastic force on atom $i$ by a directly connected
neighbor $j$ is $\vec{f}_{ij} = - {\bf K}_{ij} ( \vec{u}_i -
\vec{u}_j ) + \vec{b}_{ij} $ after linearization where the $2\times 2$
symmetric matrix ${\bf K}_{ij}=k_{ij} \hat{n}_{ij}
\hat{n}_{ij}^t $ is the modulus tensor and
$\vec{b}_{ij} = (l^0_{ij}-l_{ij}) {\bf K}_{ij} \hat{n}_{ij}$
arises from the homogeneous stress in flat films. The spring constant
$k_{ij}$ equals either $k_N$ or $k_{NN}$ for tangential or diagonal
connection respectively. The unit column vector $\hat{n}_{ij}$ points
from the unstrained lattice position of atom $j$ towards that of atom
$i$ and $t$ denotes transpose.  Furthermore, $l^0_{ij}$ and $l_{ij}$
are respectively the natural and homogeneously strained spring lengths
which follow easily from $a_s$ and $\epsilon$.  Mechanical
equilibrium requires $\sum_j \vec{f}_{ij}=0$ for each atom $i$. This
leads to a large set of equations coupling the $\vec{u}_i$ of $all$ 
of the atoms.
The solution then gives the elastic energy stored in every spring and hence
$E_s$.

We now introduce a Green's function approach for calculating $E_s$
efficiently requiring the explicit consideration of $only$ the surface
atoms. It is a lattice analogue of boundary integral methods and is
superior to boundary element techniques for our intrinsically discrete
problem. We first derive the exact formalism.  Figure 1 shows an
example of a small lattice of atoms (solid circles).  As a
mathematical construct, we extend the lattice by adding $ghost$ atoms
(open circles) with similar elastic properties.  Unphysical couplings
are hence introduced but can be exactly cancelled by applying 
{\it external} forces $\vec{f}^e_j$ and $\vec{f}^e_{j'}$ to every real
surface atom $j$ and ghost surface atom $j'$ respectively with
\begin{eqnarray}
\label{freal2}
\vec{f}^e_j &=& \sum_{j'}({\bf K}_{jj'}\vec{u}_j -\vec{b}_{jj'})\\
\label{fghost2}
\vec{f}^e_{j'} &=& - \sum_{j} {\bf K}_{jj'} \vec{u}_j
\end{eqnarray}
The summation in Eq. (\ref{freal2}) is over each ghost atom $j'$
connected directly to the real atom $j$ and it is analogous in
Eq. (\ref{fghost2}).
It is easy to see that the real atoms are then exactly decoupled from
the ghost atoms which are held precisely at their homogeneously strained
positions. 

\begin{figure}[htp]
%{\centering \epsfxsize 0.65\columnwidth \epsfbox{exforce.eps} ~\\}
\centerline{\epsfxsize 0.65\columnwidth \epsfbox{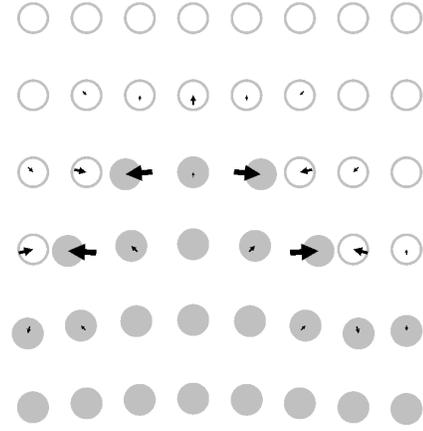}}
\vskip 2ex
\caption{
  \label{fig:exforce1}
  Ghost atoms (open circles) are added on top of the real atoms
  (solid circles) forming an extended lattice. Unphysical couplings
  are exactly cancelled by  external forces (arrows) applied to
  the real and ghost surface atoms. 
}
\end{figure}

A lattice Green's function is then applied to express the displacement,
$\vec{u}_i$, of every real surface atom $i$ under the influence of the
external forces:
\begin{equation}
\label{uife}
\vec{u}_i = \sum_j {\bf G}_{ij} \vec{f}^e_j
+ \sum_{j'} {\bf G}_{ij'} \vec{f}_{j'}^e
\end{equation}
Note that the Green's function ${\bf G}$ is
defined for the extended lattice and is {\it independent of the film
morphology.} It can thus be computed numerically once prior to the start
of the simulation. It should not be confused with the
half-plane Green's function which provides simpler but only 
approximate results \cite{Tersoff_LeGoues,Meixner}.  Combining Eqs.
(\ref{freal2}-\ref{uife}), we arrive at the reduced set of equations
\begin{equation}
\label{ui_new}
\vec{u}_i = 
\sum_{jj'} 
[({\bf G}_{ij} - {\bf G}_{ij'})
{\bf K}_{jj'} \vec{u}_j - {\bf G}_{ij} \vec{b}_{jj'}]
\end{equation}
coupling only the real surface atoms where the sum is over all pairs
of directly connected real and ghost surface atoms $j$ and $j'$
respectively. The solution of Eq. (\ref{ui_new}) gives $\vec{u}_i$.

The elastic energy $E_s$ can then be calculated directly from
\begin{equation}
\label{Es}
E_s = E_s^{0} - \frac{1}{2}\sum_{jj'} \vec{b}_{jj'} \cdot \vec{u}_j  
\end{equation}
which is derived from a simple consideration of virtual work. Here, the sum is
defined similarly to that in Eq. (\ref{ui_new}) and $E_s^{0}$ equals
the unrelaxed value of $E_s$ when $\vec{u}_i\equiv 0$ which can be
straightforwardly computed.  The method summarized in Eqs.
(\ref{ui_new}) and (\ref{Es}) is exact and practical for
simulations at moderate scales.

We can go further and use a coarse-grained version of our Green's function
approach for a further boost on the computational efficiency. Finding
the strain energy $\Delta E_m$ of the atom $m$ needed in Eq.
(\ref{rate}) requires calculating the strain energy $E_s$ of the whole
lattice twice with and without the atom $m$. Certain fine details of
the surface far away are obviously unimportant and can be neglected.
Specifically, surface atoms are grouped into sets with the $I$th of
them denoted by $\Omega_I$. We neglect fluctuations within a set by
assuming identical displacement $\vec{u}_i\equiv \vec{u}_I$ for each
member $i\in \Omega_I$. Equation (\ref{ui_new}) is then
approximated as
\begin{equation}
\label{ui_coarsened}
\vec{u}_{I} 
= \sum_J \left[
   \sum_{j\in \Omega_J,j'} 
   ({\bf G}_{Ij} - {\bf G}_{Ij'})
   {\bf K}_{jj'} 
        \right] \vec{u}_J   
  - \sum_{jj'} {\bf G}_{Ij} \vec{b}_{jj'}
\end{equation}
where $G_{Ij}=G_{ij}$ with the lattice point $i$ at the centroid of
the set $\Omega_I$. Every atom within 3 columns from the atom $m$ is
not coarsened and constitutes their own single-membered set. Farther
away at $r$ columns from the atom $m$, sets contain atoms in
neighborhoods of $2r/3+1$ columns wide, a form motivated by simple
error analysis.  We have checked numerically that a smaller degree of
coarsening leads to no noticeable difference in our results.

Surface diffusion can be simulated using the hopping rates in Eq.
(\ref{rate}) as $\Delta E_m$ is now readily computable.  We adopt an
acceptance-rejection algorithm aided by tabulated values of $\Delta
E_m$ for $5^8$ sample surface configurations. Details will be
explained elsewhere.  In our main simulations, the lattice is of 1024
atoms wide following periodic boundary conditions. The substrate is
1024 monolayers (ML) thick while the extended lattice on which we
compute the Green's function includes also a film of 80ML.  Fixed
boundary conditions are applied to the top and bottom layers of the
extended lattice.

We first simulate deposition of pure Ge film (i.e. $x=1$) with misfit
$\epsilon=4$\% at temperature $T=600K$.  At very high deposition rate
$R=80$MLs$^{-1}$ [Fig. 2(a)], we observe layer-by-layer growth. At
slower deposition rate $R=8$MLs$^{-1}$ [Fig. 2(b)], the film is
initially flat but pits then develop. A detailed examination of the
morphological evolution indicates that the pits appear rather
independently and suddenly.  Once created, they are immediately
bounded by side-walls at an energetically favorable 45$^\circ$
inclination.  These features strongly support the nucleation mechanism
for their formation, noting that pits are energetically more favorable
than islands \cite{Tersoff_LeGoues}. This pit nucleation process is
similar to that in Fig.  1(d) of Ref. \cite{Orr}.  At
$R=0.8$MLs$^{-1}$ [Fig.  2(c)], islands with 45$^\circ$ side-walls
nucleate at very early stage before the film is sufficiently thick
for pit formation. The result is analogous to those in Fig 1(a) of
Ref. \cite{Orr} and also Refs.  \cite{Barabasi,Khor}.  Further
decreasing $R$ towards realistic values of order 0.01MLs$^{-1}$ leads
to similar but more widely separated islands.

\begin{figure}
%{\centering \epsfxsize 0.95\columnwidth \epsfbox{graph.depos600.eps} ~\\}
\centerline{\epsfxsize 0.95\columnwidth \epsfbox{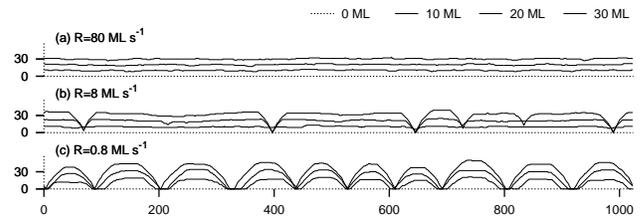}}
\vskip 2ex
\caption{
  Simulations of deposition of Ge films at $T=600K$ with substrate
  width and thickness both equal 1024$a_s \simeq$ 2780\AA. The axes
  are in unit of $a_s$} 
\end{figure}

Figure 3 shows results for deposition at misfit $\epsilon=2$\% with
$x=0.5$ at $T=1000K$.  Depending on $R$, we observe analogous
layer-by-layer growth [Fig. 3(a)], layer-by-layer growth followed by
roughening [Fig.  3(b)], and island growth [Fig.  3(c)]. However, the
islands in Figs.  3(b)-(c) emerge gradually from ripple-like
perturbations with local surface inclinations increasing steadily and
relatively synchronously in agreement with experiments
\cite{Floro,Tromp,Sutter} and ATG instability theory
\cite{Asaro,Srolovitz}. This regime has not been reported in previous
atomistic simulations mainly due to inaccuracies in accounting for the
long-range parts of the elastic interactions or the rather thin
substrates used \cite{Orr,Barabasi,Khor,Meixner}. Instead, it was
observed in continuum simulations \cite{Yang}, which however cannot
realize the nucleation mechanism.

\begin{figure}
%{\centering \epsfxsize 0.95\columnwidth \epsfbox{graph.depos1000.eps} ~\\}
\centerline{\epsfxsize 0.95\columnwidth \epsfbox{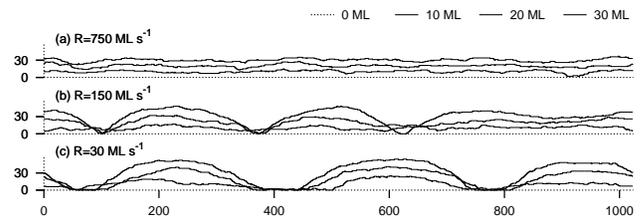}}
\vskip 2ex
\caption{
  Simulations of deposition of Si$_{0.5}$Ge$_{0.5}$ films at $T=1000K$.
} 
\end{figure}

The importance of the lattice misfit in deciding the roughening
mechanism is particularly easy to understand. The nucleation of
islands or pits occurs at a rate $R_{nucl} \sim \exp(- c \epsilon^{-4})$
with $c$ being a constant \cite{Tersoff_LeGoues} and becomes
very slow at small $\epsilon$ \cite{Sutter}.  The ATG
instability with a roughening rate $R_{inst} \sim \epsilon^{8}$
\cite{Srolovitz} then dominates.
For Figs. 2(b) and 3(b), we have chosen deposition rates close to 
the relevant roughening rates.  We then observe kinetically limited
wetting layers prior to roughening which is characteristic of
Stranski-Krastanov growth. However, the threshold thickness depends
strongly on $R$ contrary to experimental findings \cite{Politi}. A
more realistic model in the future should include 
other mechanisms such as film-substrate
interactions \cite{Tersoff91} or nonlinear elasticity \cite{Eisenberg}
which have been argued to give a more stable wetting layer.

We have also simulated annealing 
of initially flat films of 30ML at
$T=1000K$.
At this high temperature, roughening is mainly due to the ATG
instability, and we observe the development of ripples followed by
islands. Figure 4 shows the surface profiles after the islands are
fully developed.  The cusps on the surfaces \cite{Gao} are limited by
either the substrate or the local step height limit imposed in our
model. We have measured the island size $l$ from power spectra
of 11 realizations of similar surfaces.  We obtain $l \sim x
^{-1.8}$ in reasonable agreement with $x^{-2}$ from the instability
theory \cite{Asaro} but slightly different from $x^{-1}$ from
experiments \cite{Tromp,Sutter}.  The discrepancy with experiments may
be improved if the compositional dependence of film properties such as
bond energies and diffusion coefficients are properly considered.
The island sizes in general lie within the experimental range due to
our choice of the bond strength $\gamma=0.4eV$, which in fact is a
reasonable value.

\begin{figure}
%{\centering \epsfxsize 0.95\columnwidth \epsfbox{graph.depos0.eps} ~\\}
\centerline{\epsfxsize 0.95\columnwidth \epsfbox{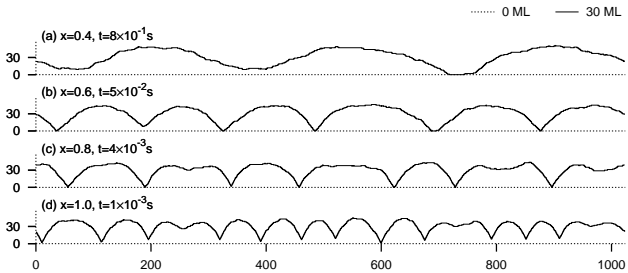}}
\vskip 2ex
\caption{
  Simulations of annealing of initially flat Si$_{1-x}$Ge$_{x}$ films
  of 30ML at $T=1000K$ for a period of time $t$.  }
\end{figure}

We have already discussed possible extensions of the model to include
film-substrate chemical interactions, nonlinear elasticity, and
composition-dependent material properties. To further enhance the
morphological resemblance with experiments, one can explore more
sophisticated forms of bond energies favoring 2D analogs of (105) and
(113) facets \cite{Floro}. 

We should note that the validity of ATG instability
theory in 3D is not clear due to a singular form of the equilibrium
surface energy in the presence of facets \cite{Politi}. Critical tests of the
theory by 3D simulations are important, and may be feasible using our
method.
For example, our 2D simulation leading to Fig. 2(b) involves about
$6\times 10^7$ hops 
%and about $3\times 10^6$ elastic energy calculations 
and ran for 18 hours on a pentium 2GHz computer.
Thus, extensions to 3D should be practical.

In conclusion, using accelerated algorithms which properly and
efficiently account for  long-range elastic interactions, we have
simulated deposition and annealing of strained heteroepitaxial layers
in 2D.
At low misfit
and high temperature, we observe ripples and subsequently gradual
island formation in consistent with the ATG instability theory. At
high misfit and low temperature, islands or pits are generated via the
nucleation pathway.
These suggest a non-trivial competition between roughening
mechanisms, although reliable quantitative determination of the
crossover conditions is beyond the scope of our model.
The ATG instability is the most promising description of island
formation in Si$_{1-x}$Ge$_{x}$ films at low, and probably also at
high lattice misfit \cite{Floro,Vailionis}. However, the nucleation
mechanism applied to high misfit regimes in certain experimental
situations has not been ruled out \cite{Kastner}. Thus the
competition of mechanisms can be important for interpreting
experimental results.
In our simulations, for deposition rates close to the relevant
roughening rate, kinetically limited wetting layers develop prior to
roughening.  At lower but more realistic deposition rates, islands
form at an early stage and are more widely separated. This may
be related to the great variation in how closely the islands are
packed under various conditions in experiments
\cite{Floro,Vailionis}.

We thank B.G. Orr for helpful comments.  CHL is supported by PolyU
grant no. 5309/01P.


\begin{references}

\bibitem{Shchukin}
% Spontaneous ordering of nanostructures on crystal surfaces
V.A. Shchukin and D. Bimberg, 
Rev. Mod. Phys. {\bf 71}, 1125 (1999).

\bibitem{Politi}
P. Politi, G. Grenet, A. Marty, A. Ponchet, and J. Villain,
Phys. Rep. {\bf 324}, 271 (2000).

%\bibitem{Mo}
%Kinetic Pathway in Stranski-Krastanov Growth of Ge on Si(001)
%Y.-W. Mo, D.E. Savage, B.S. Swartzentruber, and M.G. Lagally,
%Phys. Rev. Lett. {\bf 65}, 1020 (1990).

%\bibitem{Pimpinelli}
%A. Pimpinelli and J. Villain, Physics of crystal growth (Cambridge
%University Press, 1998).

\bibitem{Tersoff_LeGoues}
% Competing Relaxation mechanisms in strained layers
J. Tersoff and F.K. LeGoues, Phys. Rev. Lett. {\bf 72}, 3570 (1994).

\bibitem{Floro}
%Evolution of coherent islands in Si 1-x Ge x / Si(001)
J.A. Floro et al,
% E. Chason, L.B. Freund, R.D. Twesten, R.Q. Hwang, and G.A. Lucadamo, 
Phys. Rev. B {\bf 59}, 1990 (1999).

\bibitem{Vailionis}
% Pathway for the Strain-Driven Two-Dimensional to Three-Dimensional
% Transition during Growth of Ge on Si(001)
A. Vailionis et al,
% B. Cho, G. Glass, P. Desjardins, David G. Cahill, and J. E. Greene,
Phys. Rev. Lett. {\bf 85}, 3672 (2000).

\bibitem{Asaro}
R.J. Asaro and W.A. Tiller, Metall. Trans {\bf 3}, 1789 (1972);
M.A. Grinfeld, J. Nonlinear Sci. {\bf 3}, 35 (1993).

\bibitem{Srolovitz}
% On the stability of surfaces of stressed solids
D.J. Srolovitz, Acta Metall. {\bf 37}, 621 (1989);
%\bibitem{Spencer93}
%Morphological instability in epitaxially strained dislocation-free solid films: 
%linear stability theory
B. J. Spencer, P.W. Voorhees, and S.H. Davis, J. Appl. Phys. {\bf 73},
4955 (1993).


\bibitem{Tersoff01}
% Facet Growth under Stress: The limits of strained-layer stability
J. Tersoff, 
Phys. Rev. Lett. {\bf 87}, 156101 (2001).

\bibitem{Eisenberg}
% Wetting layer thickness and early evolution of epitaxially strained think films
%H.R. Eisenberg and D. Kandel, Phys. Rev. Lett. {\bf 85}, 1286 (2000).
% Thge origin and properties of the wetting layer and early evolution
%of epitaxially strained thin films
H.R. Eisenberg and D. Kandel, cond-mat 0201238 (2002).

\bibitem{Orr}
% A Model for strain-induced roughening and coherent island growth
B.G. Orr, D.A. Kessler, C.W. Snyder, and L.M. Sander,
Euro. Phys. Lett., {\bf 19}, 33 (1992).

\bibitem{Barabasi}
%Self-assembled island formation in herteropeitaxial growth
A.-L. Barab\'{a}si, Appl. Phys. Lett {\bf 70}, 2565 (1997).

\bibitem{Khor}
K.E. Khor and S. Das Sarma, Phys. Rev. B {\bf 62}, 16657 (2000).

%\bibitem{Ratsch}
% Mechanism for coherent island formation during heteroepitaxy
%C. Ratsch, P. Smilauer, D.D. Vvedensky, and A. Zangwill,
%J. Phys. I France {\bf b}, 575 (1996).  

\bibitem{Meixner}
%Self-assembled quantum dots: crossover from kinetically controlled to thermodynamically limited growth
M. Meixner et al, %E. Sch\"{o}ll, V.A. Shchukin, and D. Bimberg,
Phys. Rev. Lett. {\bf 87}, 236101 (2001)

\bibitem{Savage}
D.E. Savage et al, 
% in Germanium Silicon: Physics and materials, 
in Semiconductors and Semimetals {\bf 56}, R. Hull and J.C. Bean
Ed. (Academic Press 1999).

\bibitem{Tromp}
% Instability-driven SiGe island growth
R.M. Tromp, F.M. Ross, and M.C. Reuter,
Phys. Rev. Lett. {\bf 84} 4641 (2000).

\bibitem{Sutter}
%Nucleationless three-dimensional island formation in low-misfit heteroepitaxy
P. Sutter and M.G. Lagally, 
Phys. Rev. Lett. {\bf 84}, 4637 (2000).

\bibitem{Yang}
%Cracklike surface instabilities in stresses solids
W.H. Yang and D.J. Srolovitz,
Phys. Rev. Lett. {\bf 71}, 1593 (1993);
%\bibitem{Muller}
%model of surface instabilities induced by stress
%(Phase field method)
J. M\"{u}ller and M. Grant,
Phys. Rev. Lett. {\bf 82}, 1736 (1999).

\bibitem{Tersoff91}
% Stress-induced layer-by-layer growth of Ge on Si(100)
J. Tersoff, Phys. Rev. B {\bf 43}, 9377 (1991).

\bibitem{Gao}
%Surface roughening of heteroepitaxial thin films
H. Gao and N.D. William, Annu. Rev. mater. Sci. {\bf 29} 173 (1999).

\bibitem{Kastner}
%Kinetically self-limiting growth of Ge Islands on Si(001)
M. K\"{a}stner and B. Voigl\"{a}nder, Phys. Rev. Lett. 2745 (1999).

\end{references}
\end{document}